\documentclass[prl,twocolumn,floatfix,nobibnotes,superscriptaddress]{revtex4}
\usepackage{amsmath}
\usepackage{graphicx}
\usepackage{comment}
\usepackage[tight]{subfigure}

\newcommand{\be}{\begin{equation}}
\newcommand{\ee}{\end{equation}}

\usepackage[usenames,dvipsnames]{color}

\usepackage{color}

\begin{document}
\title{Theory of non-collinear interactions beyond Heisenberg exchange; applications to bcc Fe}
\author{A. Szilva}
\affiliation{Department of Physics and Astronomy, Division of Materials Theory, Uppsala University,
Box 516, SE-75120, Uppsala, Sweden}
\author{D. Thonig}
\affiliation{Department of Physics and Astronomy, Division of Materials Theory, Uppsala University,
Box 516, SE-75120, Uppsala, Sweden}
\author{P. F. Bessarab}
\affiliation{Science Institute of the University of Iceland, 107 Reykjavik, Iceland}
\affiliation{ITMO University, 197101 St. Petersburg, Russia}
\author{Y. O. Kvashnin}
\affiliation{Department of Physics and Astronomy, Division of Materials Theory, Uppsala University,
Box 516, SE-75120, Uppsala, Sweden}
\author{D. C. M. Rodrigues}
\affiliation{Faculdade de F\'isica, Universidade Federal do Par\' a, Bel\'em, 66075-110, Brazil}
\affiliation{Department of Physics and Astronomy, Division of Materials Theory, Uppsala University,
Box 516, SE-75120, Uppsala, Sweden}
\author{R. Cardias}
\affiliation{Faculdade de F\'isica, Universidade Federal do Par\' a, Bel\'em, 66075-110, Brazil}
\affiliation{Department of Physics and Astronomy, Division of Materials Theory, Uppsala University,
Box 516, SE-75120, Uppsala, Sweden}
\author{M. Pereiro}
\affiliation{Department of Physics and Astronomy, Division of Materials Theory, Uppsala University,
Box 516, SE-75120, Uppsala, Sweden}
\author{L. Nordstr\"{o}m}
\affiliation{Department of Physics and Astronomy, Division of Materials Theory, Uppsala University,
Box 516, SE-75120, Uppsala, Sweden}
\author{A. Bergman}
\affiliation{Maison de la Simulation, USR 3441, CEA-CNRS-INRIA-Universit\'e Paris-Sud-Universit\'e de Versailles, F-91191 Gif-sur-Yvette, France}
\affiliation{INAC-MEM, CEA, F-38000 Grenoble, France}
\author{A. B. Klautau}
\affiliation{Faculdade de F\'isica, Universidade Federal do Par\' a, Bel\'em, 66075-110, Brazil}
\author{O. Eriksson}
\affiliation{Department of Physics and Astronomy, Division of Materials Theory, Uppsala University,
Box 516, SE-75120, Uppsala, Sweden}
\affiliation{School of Science and Technology, \"{O}rebro University, SE-701 82 \"{O}rebro, Sweden}

\begin{abstract}

We show for a simple non-collinear configuration of the atomistic spins (in particular, where one spin is rotated by a finite angle in a ferromagnetic background) that the pairwise energy variation computed in terms of multiple scattering formalism cannot be fully mapped onto a bilinear Heisenberg spin model even in the lack of spin-orbit coupling. The non-Heisenberg terms induced by the spin-polarized host appear in leading orders in the expansion of the infinitesimal angle variations. However, an $E_{g}$-$T_{2g}$ symmetry analysis based on the orbital decomposition of the exchange parameters in bcc Fe leads to the conclusion that the nearest neighbor exchange parameters related to the $T_{2g}$ orbitals are $essentially$ Heisenberg-like: they do not depend on the spin configuration, \textit{and} can in this case be mapped onto a Heisenberg spin model even in extreme non-collinear cases. 
 
\end{abstract}
\pacs{later}

\maketitle

\section{I. Introduction}

The microscopic origin of the exchange interactions in ferromagnetic bcc Fe (as well as in metallic magnets in general) is a part of ongoing scientific discussions \cite{patrik, yaro,bottcher}, in spite of the fact that iron is probably the best-known magnet. Especially at finite temperature when the atomistic spin moments deviate from a collinear and uniform direction, i.e., a global magnetization axis cannot be easily identified, the dependence of the interatomic exchange parameters on the underlying spin configuration could become more significant \cite{bottcher}. The lack of a global quantization axis requires the use of a non-collinear framework for calculation of the exchange parameters which are crucial for the interpretation of the experimental observations \cite{kubler, borje}. Based on the classical Heisenberg model,
\begin{equation}
\mathcal{H}=-\sum_{i\neq j}J_{ij}\vec{e}
_{i} \cdot \vec{e}_{j}\;,
\label{BL}
\end{equation}%
where $\vec{e}_{i}$ $\left(\vec{e}_{j}\right)$  is a unit vector pointing in the direction of the atomic moment at site $i$ ($j$), and $J_{ij}$ stands for the exchange coupling parameter between the magnetic moments, the critical temperature and the magnon excitation spectra of iron at low temperatures can be well described by \textit{ab initio} calculations \cite{bottcher, borje}. Note that sum in Eq. (\ref{BL}) avoids double counting.

Although a formula for the exchange coupling, $J_{ij}$, in case of collinear arrangement has been known for a long time, due to the seminal work of Lichtenstein, Katsnelson, Antropov, Gubanov (LKAG) \cite{oldlicht}, even for relativistic \cite{udvardi} and correlated systems \cite{newlicht}, a counterpart mapping onto a spin Hamiltonian for non-collinear arrangement is non-trivial. Similar to the LKAG derivation, a derivation was found in Ref. \cite{szilva} for the pairwise energy variation based on magnetic force theorem \cite{mac, meth} by allowing the presence of a non-collinear underlying spin configuration. The pairwise energy variation term, $\delta E_{ij}^{two}$, emerges for the case when two atomistic spins are infinitesimally rotated at two different sites \textit{at the same time}, and its further analysis is in the scope of this paper. Note that the total energy variation can be written as $\delta E_{ij}=\delta E_{i}^{one}+\delta E_{j}^{one}+\delta E_{ij}^{two}$ where the one site energy variation $\delta E_{i}^{one}$ ($\delta E_{j}^{one}$) takes into account the interaction between the spin at site $i$ ($j$) and the environment formed by the other spins (except of the spins sitting at site $i$ and $j$). A system is equilibrium when $\delta E_{i}^{one}$ does not have finite contribution of leading order in the infinitesimal rotation angle $\delta \theta_{i}$. One can still find non-collinear systems also in equilibrium as the flat spin spirals \cite{flat}. By using the non-collinear approach for bcc Fe as well as for Fe overlayers on Ir(001), the magnon softening observed at room temperature in neutron scattering experiments \cite{Lynn} was explained \cite{szilva, debora}. In this article, we show that in a general, non-equilibrium, non-collinear case, an anisotropic type term is found that cannot be mapped onto a Heisenberg model given by Eq. (\ref{BL}) even in the lack of spin-orbit coupling. We show, however, that an $E_{g}$-$T_{2g}$ symmetry analysis based on the orbital decomposition of the $J_{ij}$ leads to a similar conclusion for the origin of Heisenberg and non-Heisenberg terms as was published in Ref. \cite{yaro} within the LKAG approach. 

\begin{figure}[b]
\begin{center}
\begin{tabular}
[c]{c}%
\includegraphics[width=0.5\textwidth, bb= -150 -10 580 300 ]{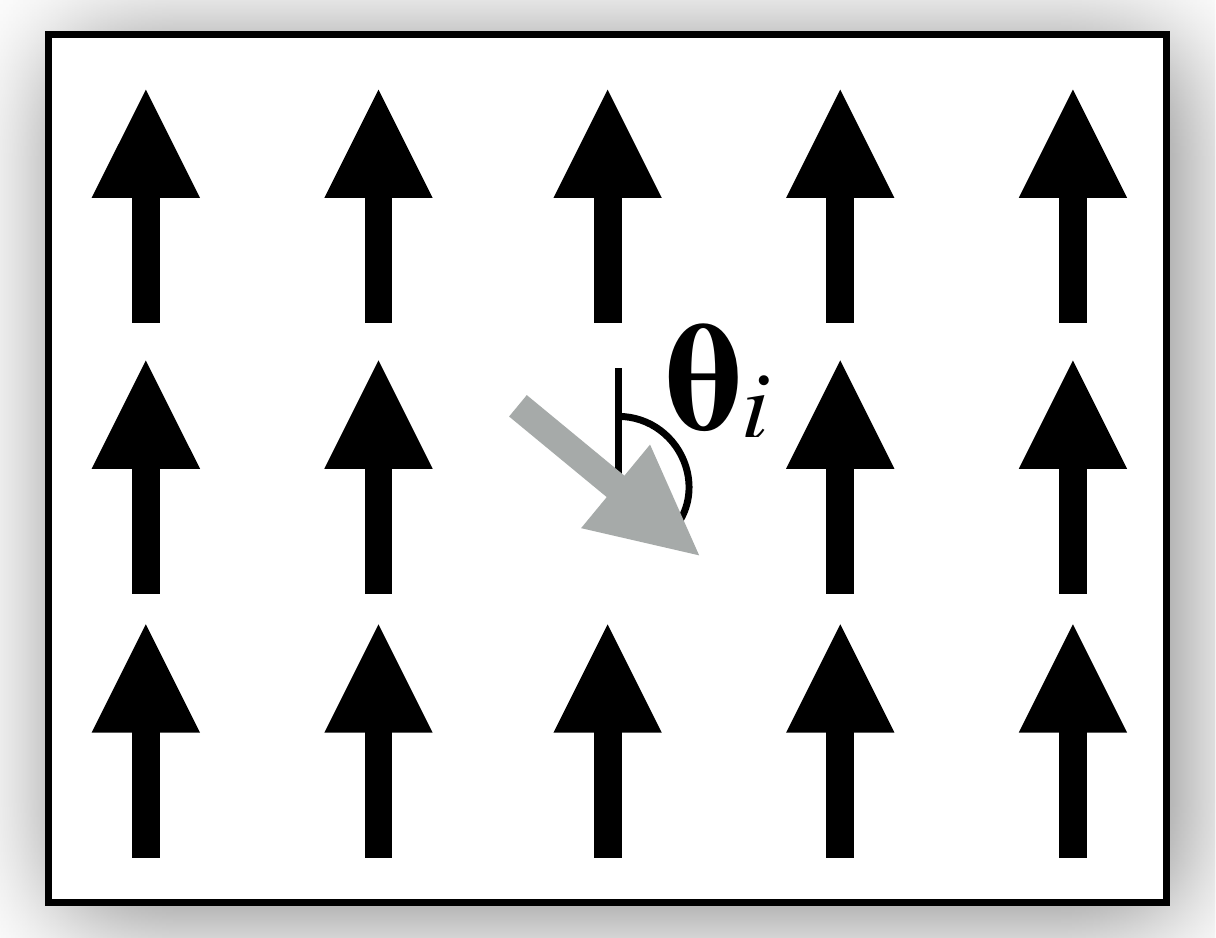}
\end{tabular}
\end{center}
\par
\vskip -0.5cm  \caption{(color online) Schematic representation of the geometry of the \textit{single spin rotation system}. As a new system of reference for the infinitesimal two-site spin rotations: an atomistic spin is rotated by a finite $\theta_{i}$ angle at site $i$ on a lattice when every other spin forms a ferromagnetic background.}%
\label{scheme}%
\end{figure}

Higher-order exchange interactions are known to emerge in the absence of spin-orbit coupling. Recently the microscopic theory of the magnetic interactions including four-spin exchange has been successfully applied to simulate the magnetic phase diagram of heavy rare earth elements \cite{rareearth}. In case of bcc Fe, higher-order (biquadratic) exchange interactions have to be also taken into account in the spin Hamiltonian \cite{lounis,Y20}. It can be shown that the non-collinear pairwise energy variation formula recovers the pairwise energy formula published in Ref. \cite{lounis} for a collinear case by keeping the higher (fourth) order terms in the infinitesimal angle variation, i.e., an anisotropic term is present even in the collinear limit, which was found to be numerically significant in bcc Fe \cite{szilva}. Here, we will show that in a non-equilibrium case, when one spin is rotated by a finite angle in a ferromagnetic background (\textit{single spin rotation} shown in Fig. \ref{scheme}), one can find contributing terms in the non-collinear $\delta E_{ij}^{two}$ that are \textit{essentially} non-Heisenberg of \textit{the leading} (second) order in the infinitesimal angle variation: terms can be found that cannot be mapped onto Eq. (\ref{BL}). This is a different case than a Heisenberg model with spin-configuration dependent $J_{ij}$'s as was defined in Ref. \cite{yaro}, where $J_{ij}$ in Eq. (\ref{BL}) was referred to as non-Heisenberg parameter when it (significantly) depended on the spin configuration. However, in this article, we numerically calculate the implicit configuration dependence of the parameters that are needed to determine the pairwise energy variation $\delta E_{ij}^{two}$. It should also be noted that it is possible to do a spin-cluster expansion for a description in terms of spin models \cite{deak}.

The article is structured as follows. In Section II, we will outline the energy variation formula of the pairwise derived in Ref. \cite{szilva} for a general, non-collinear spin arrangement, and apply for the system of \textit{single spin rotation}. Then, in Section III, we summarize the technical details of the density functional theory (DFT) calculations based on multiple scattering formalism (MSF) \cite{gyorffy}. The results of the article will be presented in Section IV, while Section V will summarize the main conclusions.

\section{II. Theoretical background}

The fundamental equation of a scalar relativistic MSF is given as \cite{gyorffy}
\begin{equation}
\left( \tau _{ij}^{-1}\right) _{L\sigma ,L^{\prime }\sigma ^{\prime
}}=p_{iL\sigma \sigma ^{\prime }}\delta _{ij}\delta _{LL^{\prime
}}-G^{0}_{ij,LL^{\prime }}\delta _{\sigma \sigma ^{\prime }}\;,
\label{FEMSTT}
\end{equation}%
where $\tau _{ij}$ stands for the scattering path operator (SPO), $\mathbf{p}_{i}=\mathbf{t}_{i}^{-1}$, denotes the inverse of single site scattering operator (ISO). In Eq. (\ref{FEMSTT}) $L=\left(l,m \right)$ stands for the angular momentum and magnetic quantum numbers, $\sigma$ refers to the spin-index, ${G}_{ij}^{0}$ is the free (or bare) structure constant and indices $i$ and $j$ refer to the considered lattice sites. ${G}_{ij}^{0}$ is calculated from the Hamiltonian of the free particle, hence it is spin-independent. Later on in our presentation, we omit the orbital and spin indices, and the boldface notation stands for quantities in both spin and orbital spaces ($18 \times 18$ matrices in spd basis) while the lack of boldface refers to quantities defined only in the orbital space ($9 \times 9$ matrices in spd basis). We introduce a general notation for the single site scattering operator in a non-collinear framework as
\begin{equation}
\mathbf{t}_{i}=t_{i}^{0}I_{2}+t_{i}\vec{e}_{i} \cdot \vec{\sigma} \;,
\label{tdef}
\end{equation}%
where the unit vector $\vec{e}_{i}$ refers to the magnetic spin moment at site $i$ (as it was already defined in the Introduction), and can be written as
\begin{equation}
\vec{e}_{i}= \left(\sin(\theta_{i}) \cos(\phi_{i}), \sin(\theta_{i}) \sin(\phi_{i}), \cos(\theta_{i}) \right)
\end{equation}%
where $\theta_{i}$ and $\phi_{i}$ are the polar and azimuthal angles, respectively.  $\vec{\sigma}$ is the vector formed by Pauli-matrices, $I_{2}$ is the unit matrix in spin space, $t_{i}^{0}$ denotes the non-magnetic (charge) part, and ${t}_{i}$ stands for the magnetic (spin) part of the single site scattering operator. Note that the single site scattering operator, $\mathbf{t}_{i}$, depends on the energy, $\varepsilon$.

For the ISO, one can introduce the same notation as for the $\mathbf{t}_{i}$ in Eq. (\ref{tdef}) as follows,
\begin{equation}
\mathbf{p}_{i}=p_{i}^{0}I_{2}+p_{i}\vec{e}_{i} \cdot \vec{\sigma} \;.
\label{pdef}
\end{equation}
Later we will need to deal with the variation of the ISO under a small rotation that can be written as
\begin{equation}
\delta \mathbf{p}_{i}=p_{i}\delta \vec{e}_{i} \cdot \vec{\sigma}\;, \label{deltapii}
\end{equation}%
where $\delta\vec{e}_{i}$ stands for the deviation of a spin moment after an infinitesimal rotation at site $i$. Finally, the SPO has a structure as
\begin{equation}
\mathbf{\tau} _{ij}=T_{ij}^{0}I_{2}+\vec{T}_{ij} \cdot \vec{\sigma}\;,
\label{deftau3}
\end{equation}%
where $T_{ij}^{0}$ denotes the charge while $\vec{T}_{ij}=\left(T^{x}_{ij}, T^{y}_{ij}, T^{z}_{ij} \right)$ stands for the spin part of the SPO. In collinear limit  $\vec{T}_{ij}$ is reduced to $\left(0, 0, T^{z}_{ij} \right)$, and the component of SPO in the up and down spin channels can be defined as $T_{ij}^{\uparrow}=T_{ij}^{0}+T_{ij}^{z}$ while $T_{ij}^{\downarrow}=T_{ij}^{0}-T_{ij}^{z}$.

\begin{figure}[b]
\begin{center}
\begin{tabular}
[c]{c}%
\includegraphics[width=0.5\textwidth, bb= 20 0 575 380 ]{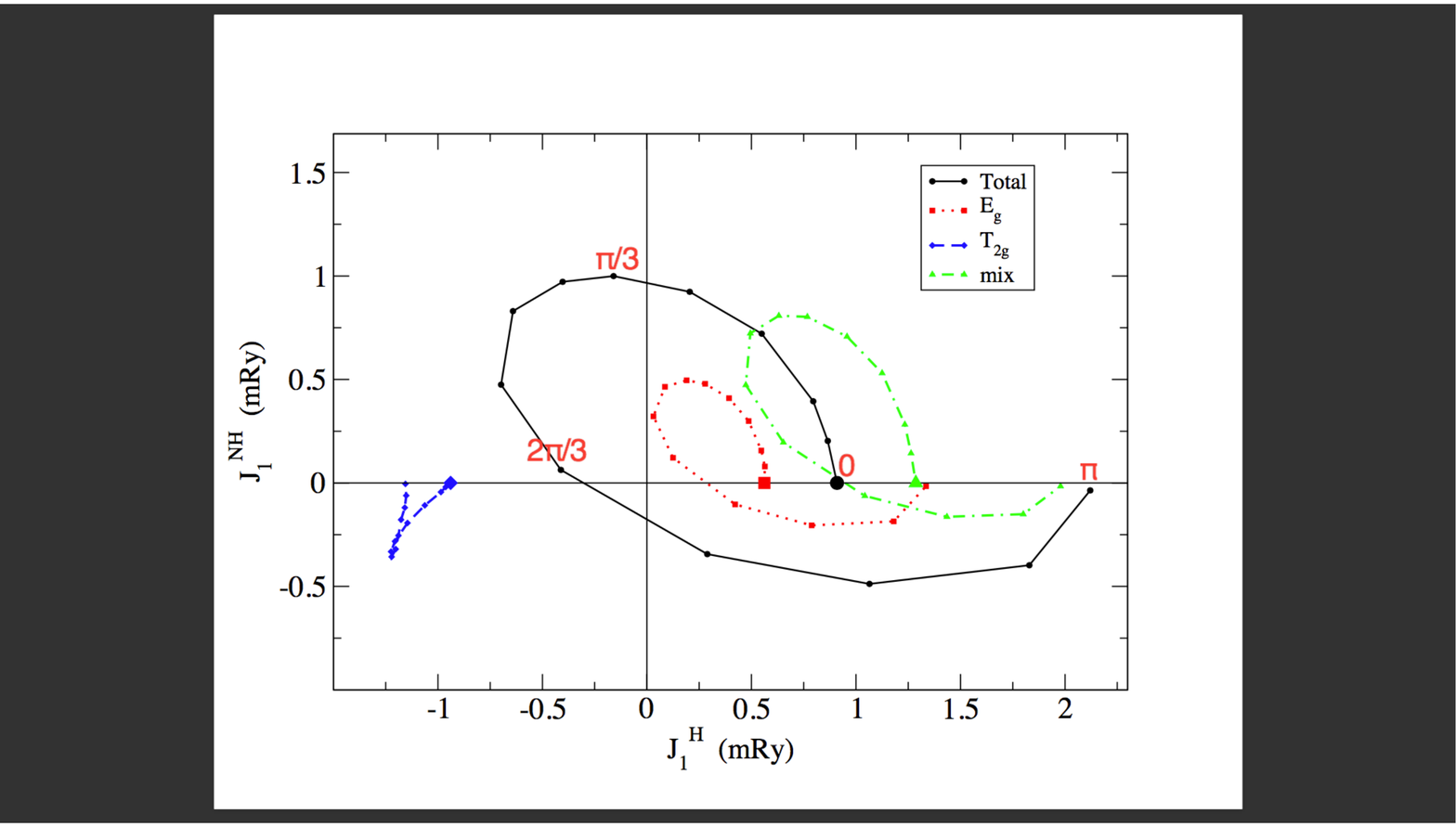}
\end{tabular}
\end{center}
\par
\vskip -0.5cm  \caption{(color online)
The black line shows the evolution of the non-collinear interatomic exchange coupling constants for a first nearest neighbour pair in bcc Fe single spin rotation, see Eqs. (\ref{H}) and Eq. (\ref{NH}). The red, blue and green lines stand for its symmetry decomposition in the d-channel defined by Eq. (\ref{decomp}). The parameters in the ferromagnetic collinear limit when $\theta_{i}=0$ are denoted by bigger symbols. For the non-decomposed (black) curve, the $\theta_{i}=\pi/3$, $\theta_{i}=2\pi/3$ and $\theta_{i}=\pi$ are also noted.}%
\label{J1}%
\end{figure}

\begin{figure}[b]
\begin{center}
\begin{tabular}
[c]{c}%
\includegraphics[width=0.5\textwidth, bb= 50 0 1300 870 ]{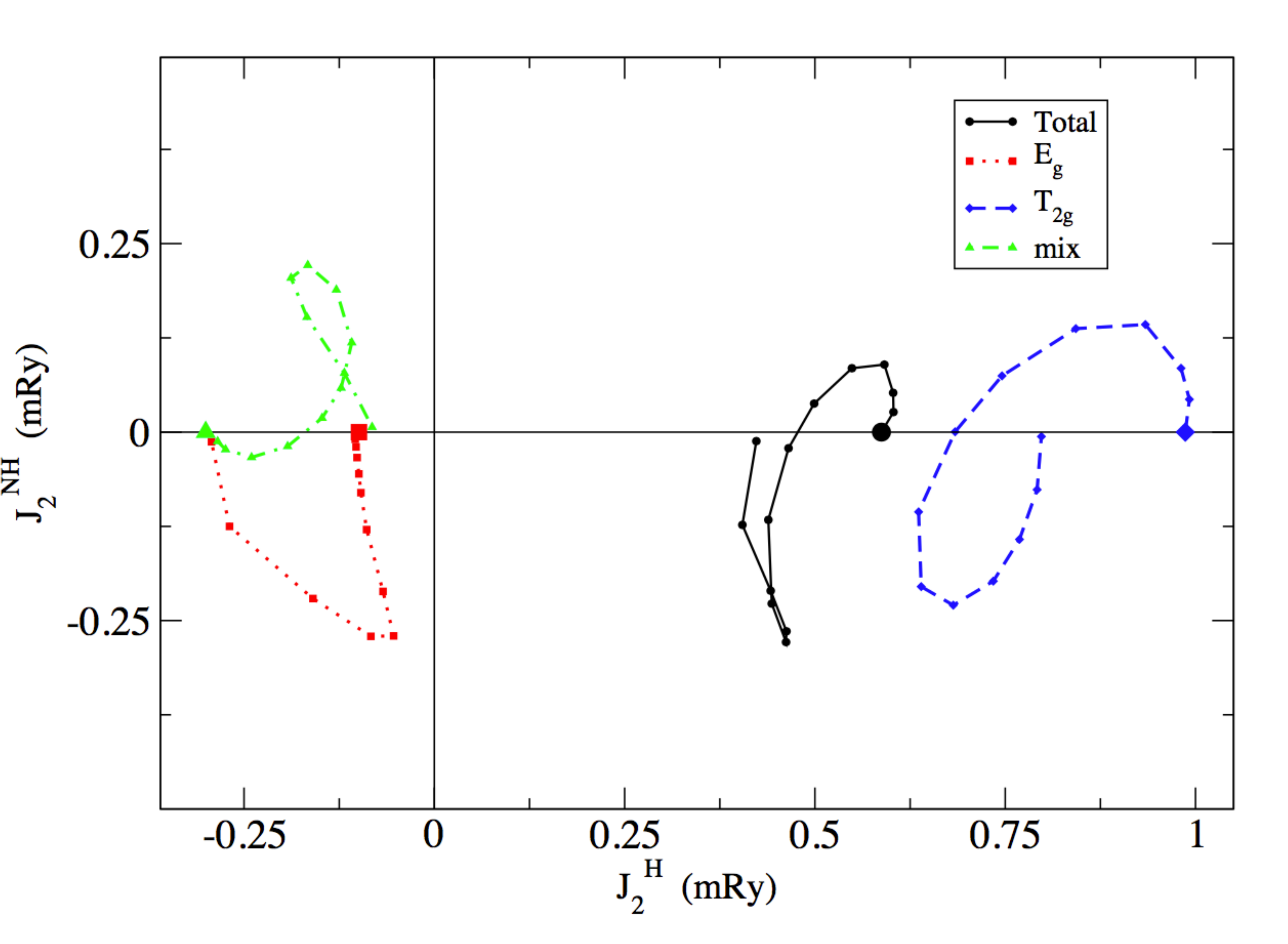}
\end{tabular}
\end{center}
\par
\vskip -0.5cm  \caption{(color online)
The black line shows the evolution of the non-collinear interatomic exchange coupling constants for a \textit{second} nearest neighbour pair in bcc Fe single spin rotation, see Eqs. (\ref{H}) and Eq. (\ref{NH}). The red, blue and green lines stand for its symmetry decomposition in the d-channel defined by Eq. (\ref{decomp}). The parameters in the ferromagnetic collinear limit when $\theta_{i}=0$ are denoted by bigger symbols.}%
\label{J2}%
\end{figure}

\begin{figure}[b]
\begin{center}
\begin{tabular}
[c]{c}%
\includegraphics[width=0.5\textwidth, bb= 40 0 1490 990 ]{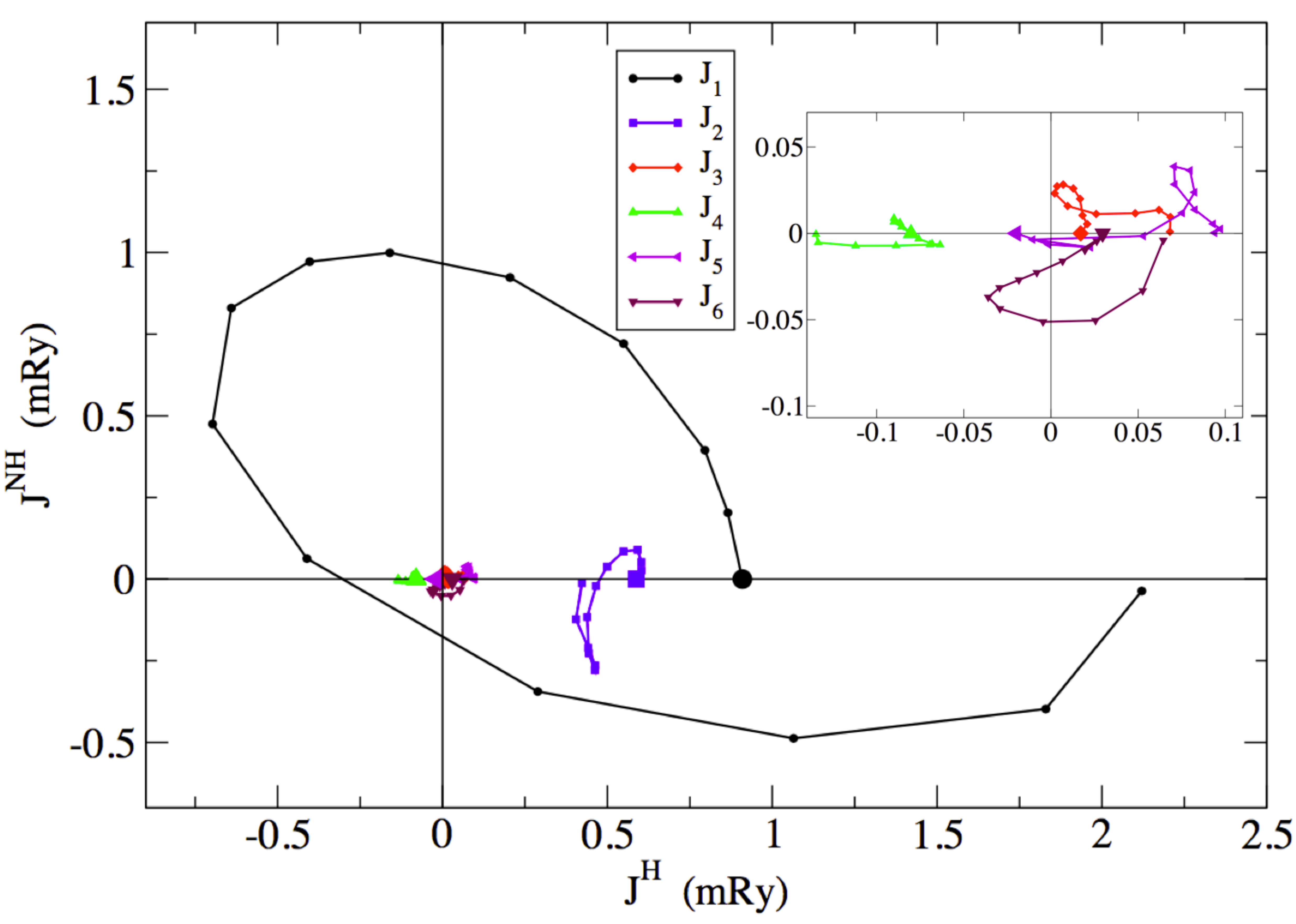}
\end{tabular}
\end{center}
\par
\vskip -0.5cm  \caption{(color online)
The calculated $J^{H}$ and $J^{NH}$ parameters for the first six nearest neighbors in bcc Fe in the (total) d sector in case of single spin rotation. The parameters in the ferromagnetic collinear limit when $\theta_{i}=0$ are denoted by bigger symbols. Inset:  The same parameters for the third, fourth, fifth and sixth neigbours.}%
\label{NN}%
\end{figure}

So far, we have defined the quantities required to calculate the pairwise total energy variation in a non-collinear framework. This can be obtained as the variation of the integrated density of states times energy \cite{meth, lloydd}. Due to the Lloyd formula \cite{lloydd}, it can be written that
\begin{equation}
\delta E_{ij}^{two} = -\frac{1}{\pi }\int\limits_{-\infty
}^{\varepsilon _{F}}d\varepsilon \operatorname{Im}Tr_{\sigma L}\left( \delta
\mathbf{p}_{i}\mathbf{\tau} _{ij}\delta \mathbf{p}_{j}\mathbf{\tau} _{ji}\right) \;,  \label{deltae}
\end{equation}%
where $\varepsilon _{F}$ stands for the Fermi energy. Inserting Eqs. (\ref{deltapii}) and (\ref{deftau3}) into Eq. (\ref{deltae}), and introducing the matrix
\begin{equation}
A_{ij}^{\alpha \beta }=\frac{1}{\pi }\int\limits_{-\infty }^{\varepsilon_{F}}d\varepsilon \operatorname{Im} Tr_{L}\left( p_{i}T_{ij}^{\alpha
}p_{j}T_{ji}^{\beta }\right)   \label{AImdeff}
\end{equation}%
where indices $\alpha $ and $\beta $ run over \footnote{In the collinear LKAG limit $\alpha $ and $\beta $ run over $\uparrow$ and $\downarrow$, referring to the up and down spin channels.} $0$, $x$, $y$ or $z$, we get that 
\begin{equation}
\delta E_{ij}^{two}=-2\left( A_{ij}^{00}-\sum_{\mu}A_{ij}^{\mu \mu
}\right) \delta \vec{e}_{i} \cdot \delta \vec{e}
_{j}-4\sum_{\mu ,\nu } A_{ij}^{\mu \nu } \delta e_{i}^{\mu }\delta
e_{j}^{\nu }\;, \label{gen22}
\end{equation}%
where $\mu$ and $\nu$ run over $x$, $y$ and $z$. In Eq. (\ref{gen22}), we have repeated the derivation of Ref. \cite{szilva} for the non-collinear pairwise energy variation by introducing the quantities we will need to present our results.

\section{III. Details of the calculations}

The calculations were performed with the use of standard DFT techniques by means of real-space linear muffin-tin orbital method within the atomic sphere approximation (RS-LMTO-ASA) \cite{Y30,Y31}. We employed the standard local spin density approximation (LSDA) for the exchange-correlation energy throughout the study. 

First we have calculated self-consistently the electronic structure of system with 8393 Fe atoms arranged into a bcc lattice structure where its inner (core) part can be considered bulk-like. The interatomic distance was set as 2.861 \r{A} while the Wigner-Seitz radius was as 1.409 \r{A}. Then we embedded 9 Fe impurity atoms (one atom and its nearest neighbours) into the Fe bulk host making it possible to change the direction of the atomistic spins on those atoms, and to calculate the electronic structure self-consistently for every $\theta_{i}$ angle which was set at site $i$ while the spin direction on other atoms was kept collinear, ferromagnetic \footnote{In this article we modify the spin direction only for the "centre" atom. Also note that the exchange parameters can be calculated for further neighbour pairs too.}.

\section{IV. Results}

\begin{figure}[b]
\begin{center}
\begin{tabular}
[c]{c}%
\includegraphics[width=0.5\textwidth, bb= 30 20 1250 850 ]{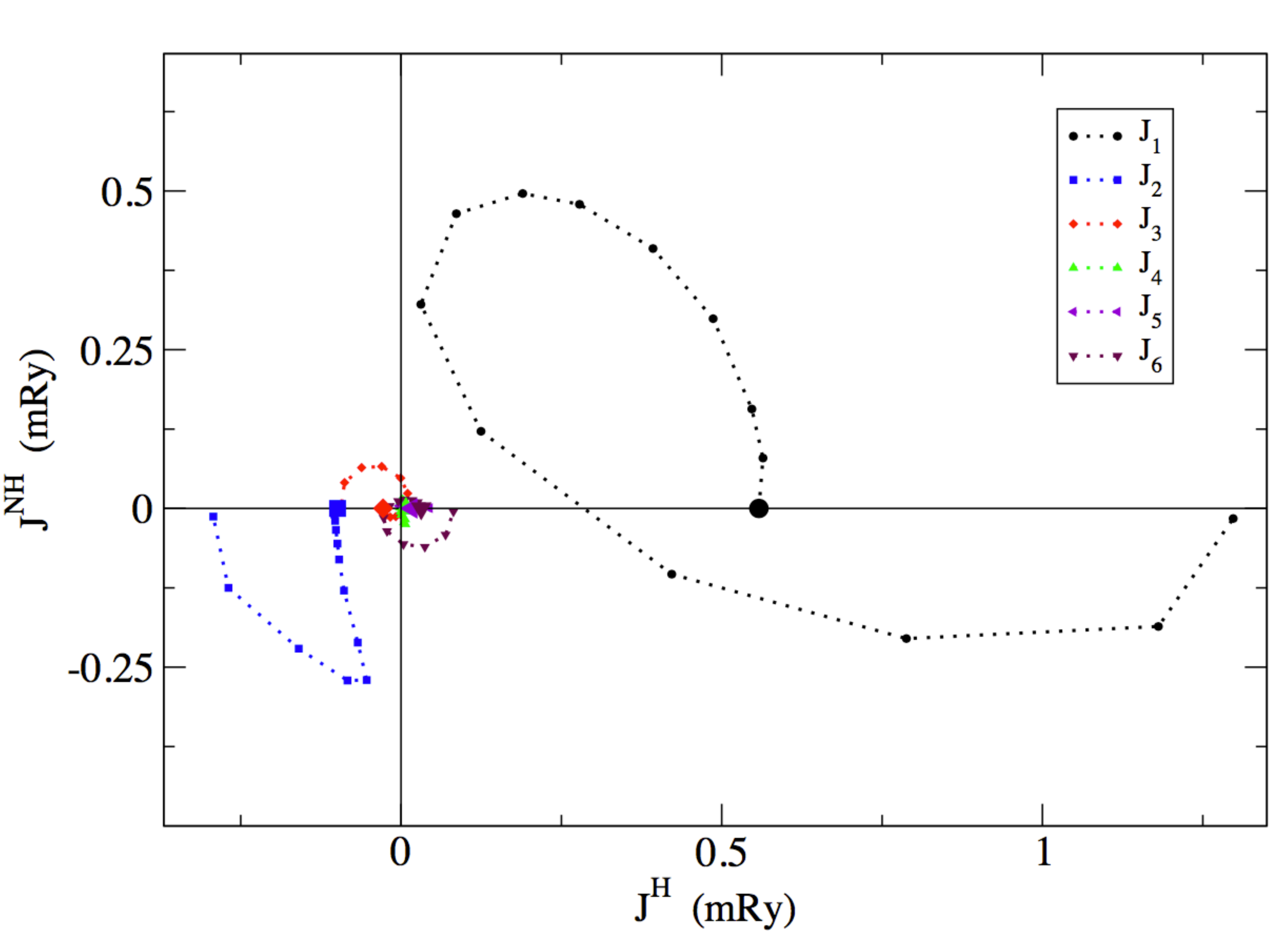}
\end{tabular}
\end{center}
\par
\vskip -0.5cm  \caption{(color online)
The calculated $J^{H}$ and $J^{NH}$ parameters for the first six nearest neighbors in bcc Fe in the $E_{g}$ sector in case of single spin rotation. The parameters in the ferromagnetic collinear limit when $\theta_{i}=0$ are denoted by bigger symbols.}%
\label{NNE}%
\end{figure}

From a Heisenberg spin model as the one introduced in Eq. (\ref{BL}), one gets for the pairwise energy variation, in general, that
\begin{equation}
\delta E_{ij}^{two-H}=-2 J_{ij} \delta \vec{e}_{i} \cdot \delta \vec{e}
_{j}\;, \label{HH1}
\end{equation}%
which is reduced to 
\begin{equation}
\delta E_{ij}^{two-H}= 2 J_{ij} \delta \theta_{i} \delta \theta_{j}  \label{HH2}
\end{equation}%
in collinear limit when only the leading term is kept by inserting the spin variations $\delta \vec{e}_{i}=\left (\delta \theta_{i}, 0, 0 \right)$ and $\delta \vec{e}_{j}=\left (-\delta \theta_{j}, 0, 0 \right)$. The symbol $H$ in the expression $\delta E_{ij}^{two-H}$ refers to the fact that it is derived from the Heisenberg model Eq. (\ref{BL}). In this case every spin points to $z$ direction in a global coordinate system, and a spin at site $i$, and an other one at site $j$ are rotated by the angle $\delta \theta_{i}$ and $\delta \theta_{j}$, respectively. It was shown in Ref. \cite{szilva} that in this collinear case $J_{ij}=A^{00}_{ij}-A^{zz}_{ij} = A^{\uparrow \downarrow}_{ij}$. The LKAG formula \cite{oldlicht} can be derived in collinear limit where the $T_{ij}^{\uparrow}$ and $T_{ij}^{\downarrow}$ can be defined as was shown in Section II.

\begin{figure}[b]
\begin{center}
\begin{tabular}
[c]{c}%
\includegraphics[width=0.5\textwidth, bb= 70 0 1500 1000 ]{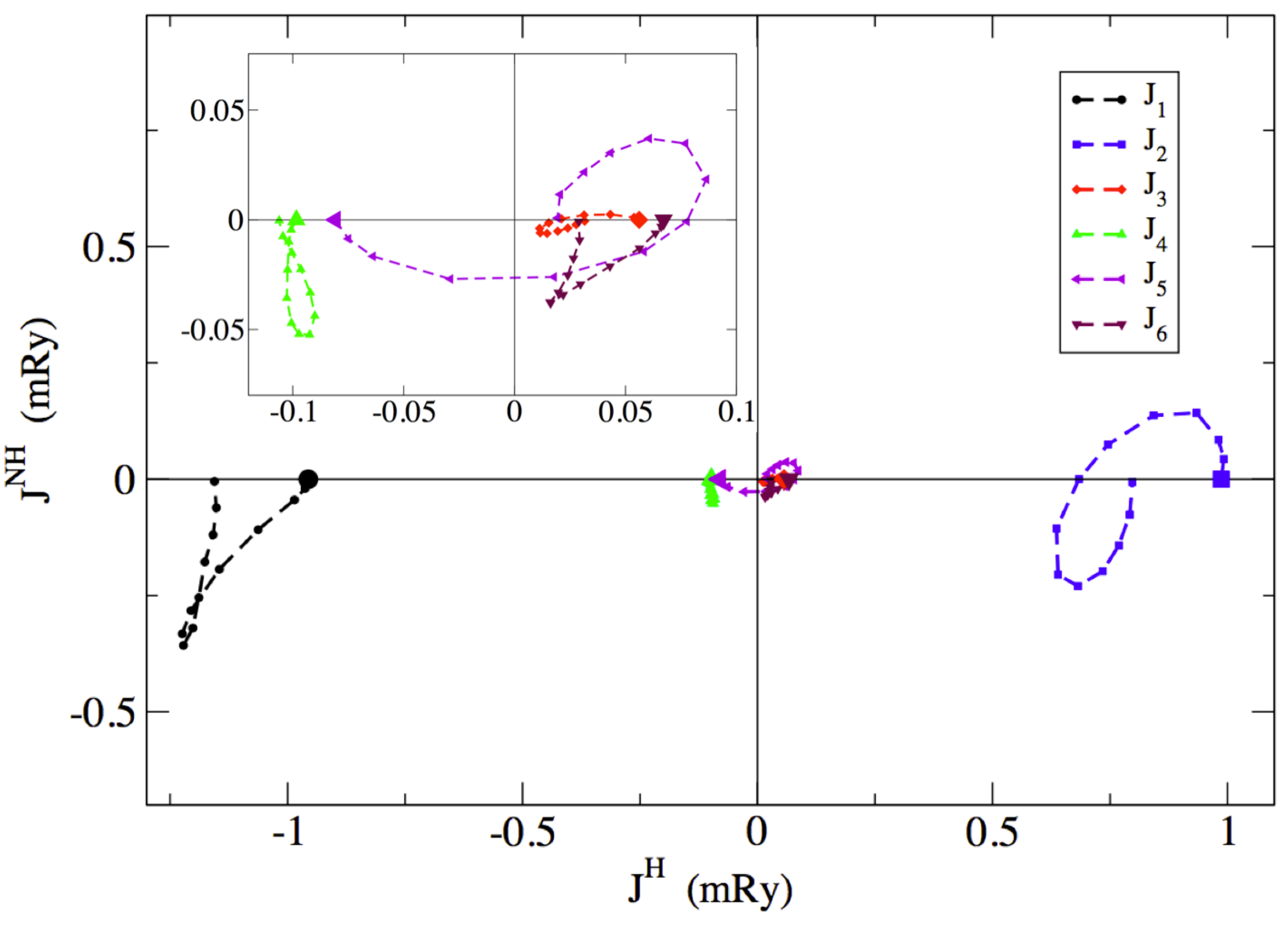}
\end{tabular}
\end{center}
\par
\vskip -0.5cm  \caption{(color online)
The calculated $J^{H}$ and $J^{NH}$ parameters for the first six nearest neighbors in bcc Fe in the $T_{2}$ sector in case of single spin rotation. The parameters in the ferromagnetic collinear limit when $\theta_{i}=0$ are denoted by bigger symbols. Inset:  The same parameters for the third, fourth, fifth and sixth neigbours.}%
\label{NNT2}%
\end{figure}

\begin{figure}[b]
\begin{center}
\begin{tabular}
[c]{c}%
\includegraphics[width=0.5\textwidth, bb= 30 0 1250 850 ]{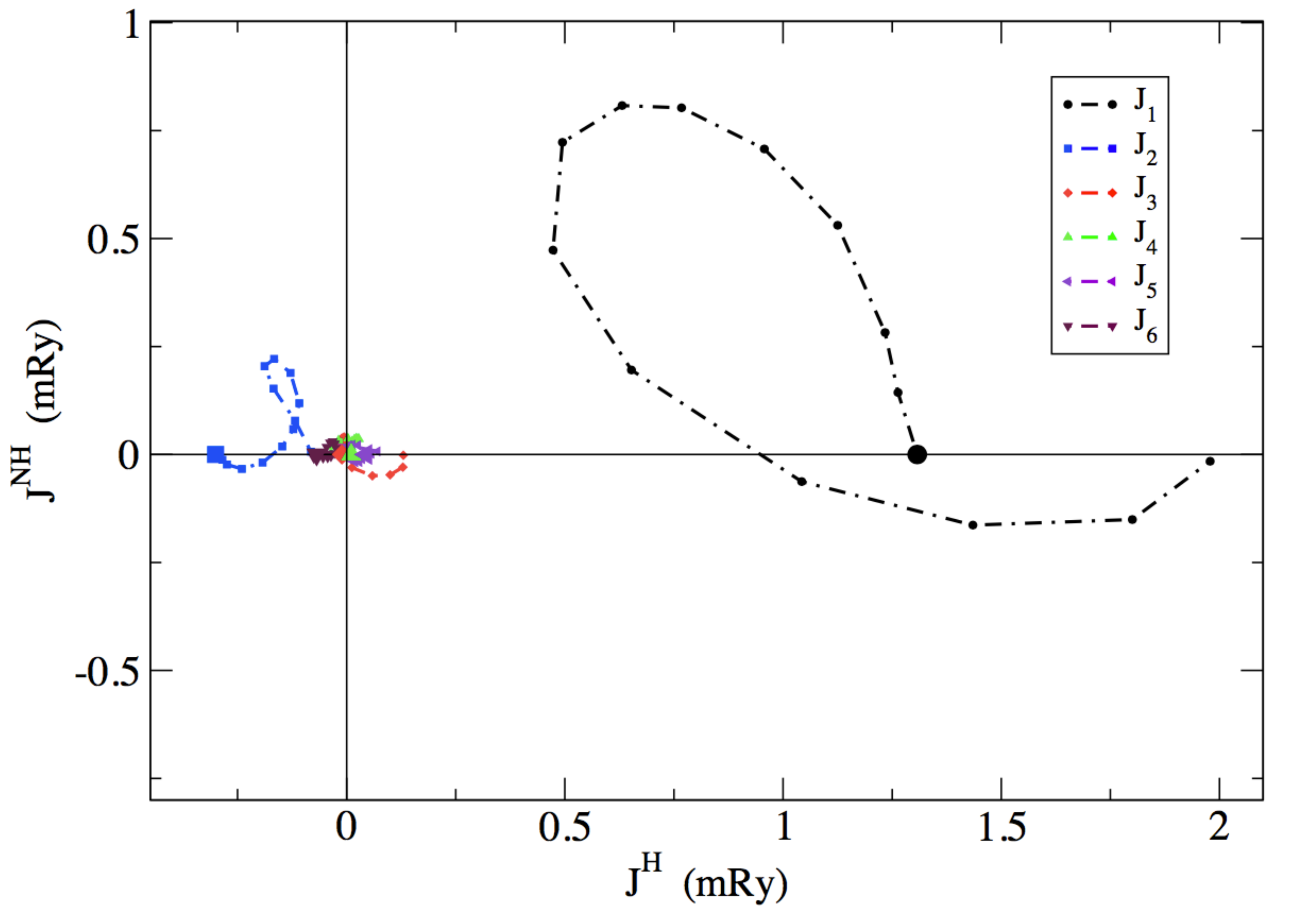}
\end{tabular}
\end{center}
\par
\vskip -0.5cm  \caption{(color online)
The calculated $J^{H}$ and $J^{NH}$ parameters for the first six nearest neighbors in bcc Fe in the mixed sector in case of single spin rotation. The parameters in the ferromagnetic collinear limit when $\theta_{i}=0$ are denoted by bigger symbols.}%
\label{NNmix}%
\end{figure}

Next, we consider the case when $\vec{e}_{i}=\left (\sin \theta_{i}, 0,\cos \theta_{i} \right)$, i.e., the spin at site $i$ is rotated in the $xz$ plane by $\theta_{i}$, see Fig. \ref{scheme}, while every other spin forms a ferromagnetic background, i.e., $\vec{e}_{j}=\left (0, 0, 1 \right)$. This provides a new reference frame for the infinitesimal rotations, and will be referred to as the \textit{single spin rotation}. It can then be shown that $\delta \vec{e}_{i}=\left (\cos \theta_{i}, 0,-\sin \theta_{i} \right) \delta \theta_{i}$ when the spin at site $i$ is rotated now by $\delta \theta_{i}$ while $\delta \vec{e}_{j}$ remains $\left (-\delta \theta_{j}, 0, 0 \right)$. Inserting $\delta \vec{e}_{i}$ and $\delta \vec{e}_{j}$ into Eq. (\ref{HH1}), we get for the pairwise energy variation in terms of a Heisenberg model that
\begin{equation}
\delta E^{two-H}_{ij} =2  J_{ij} \cos \left(\theta_{i} \right)   \delta \theta_{i}  \delta \theta_{j}      \;,
\end{equation}%
i.e., we get that the pairwise energy is proportional to a $cos \left(\theta_{i} \right)$ which recovers Eq. (\ref{HH2}) in the LKAG limit. Inserting $\delta \vec{e}_{i}$ and $\delta \vec{e}_{j}$ however into Eq. (\ref{gen22}), we get for the pairwise energy variation in terms of MSF that
\begin{equation}
\delta E^{two}_{ij} =2 \left[ J^{H}_{ij} \cos \left(\theta_{i} \right) + J^{NH}_{ij} \sin  \left(\theta_{i} \right)  \right] \delta \theta_{i}  \delta \theta_{j}      \;.
\end{equation}%
where
\begin{equation}
J^{H}_{ij} = A^{00}_{ij} + A^{xx}_{ij}- A^{zz}_{ij}
\label{H}
\end{equation}
and
\begin{equation}
J^{NH}_{ij} = -2 A^{zx}_{ij} \;,
\label{NH}
\end{equation}
where the term which is proportional to a $cos \left(\theta_{i} \right)$ and $sin \left(\theta_{i} \right)$ are referred to as Heisenberg (H) term and non-Heisenberg (NH) term, respectively. It should be stressed that both the Heisenberg, $J^{H}_{ij} \left(\theta_{i} \right)$, and the non-Heisenberg,  $J^{H}_{ij} \left(\theta_{i} \right)$, parameters can implicitly depend of the angle $\theta_{i}$, too. In the case when $\theta_{i}=\pi/2$, i.e, when the Heisenberg contribution is zero because $cos \left(\pi/2 \right)=0$, $J^{NH}_{ij} $ can still be finite. In this case the system is \textit{essentially} non-Heisenberg. Otherwise, one should describe it as a Heisenberg model with a configuration dependent $J^{H}_{ij} $.

It is easy to show in a collinear case that $J^{H}_{ij}=A^{00}_{ij}-A^{zz}_{ij} = A^{\uparrow \downarrow}_{ij}$, i.e, the LKAG formula can be recovered because $A^{xx}_{ij}$ and $A^{zx}_{ij}$ in Eq. (\ref{H}) and  (\ref{NH}) vanish then. Note that we can write as a better approximation that $ \delta \vec{e}_{i} \simeq  \left( \delta \theta_{i},0, -1/2  \left( \delta \theta_{i}   \right)^{2} \right)  $ and a similar expression for $ \delta \vec{e}_{j}$. Then we recover the results for $\delta E^{two}_{ij}$ published in Ref. \cite{lounis} where the anisotropic type term was mapped onto a four-spin Hamiltonian. Note that Eq. (\ref{deltae}) should then be also extended with further terms to be correct at the level of fourth order approximation in $\delta \theta$'s. As was reported in Ref. \cite{szilva}, for the nearest neighbor pairs \footnote{We use the short notation 1, 2, 3... in the subindex for the nearest, second nearest, third nearest... neighbouring $i$-$j$ pairs.} $A^{00}_{1}=-0.23$ mRy while $A^{zz}_{1}=-1.08$ mRy. This means that the spin ($zz$) contribution dominates the charge one ($00$) in the LKAG $J_{1}=A^{\uparrow \downarrow}_{1}$ in bcc Fe which is given as $A^{\uparrow \downarrow}_{1}=0.85$ mRy.

Similar to Eq. (1) in Ref. \cite{yaro}, we can decompose the parameters in terms of cubic symmetry group representations as follows,
\begin{equation}
A^{\alpha\beta}_{ij}= A^{\alpha\beta-E_{g}}_{ij}+ A^{\alpha\beta-T_{2g}}_{ij} + A^{\alpha\beta-mix}_{ij}   \;,
\label{decomp}
\end{equation}%
where, as it was shown in Ref. \cite{ramon}, in some cases as for the second neighbor pairs the decomposition is complete, however, in general we have a finite mixed term, $A^{\alpha\beta-mix}_{i}$. For the first neighbors we find that $A^{\uparrow \downarrow - E_{g}}_{1}= 0.56$ mRy, $A^{\uparrow \downarrow - T_{2g}}_{1}= -0.99$ mRy and  $A^{\uparrow \downarrow - mix}_{1}= 1.29$ mRy. This means that there is a pronounced antiferromagnetic coupling coming from to the $T_{2g}$ orbitals ($T_{2g}$ channel) that was shown to be related to Fermi-surface mechanisms (like RKKY oscillations calculated in the asymptotic regime) by calculating the exchange parameters for further neighbour pairs in a given direction. However, in the $E_{g}$ and mixed channels the microscopic origin is mainly the double-exchange mechanism \cite{yaro} that requires the presence of the (ferromagnetic) host. Note that we calculate $A^{\alpha\beta}_{ij}$ in Eq. (\ref{AImdeff}) only for the d electrons. The contribution of the s and p channels are negligible, and the decomposition in Eq. (\ref{decomp}) becomes then exact. 

As we can see in Eq. (\ref{gen22}), the $A^{00}_{ij}$ term contributes only to the Heisenberg type term, being proportional to $\delta \vec{e}_{i} \cdot \delta \vec{e}_{j}$. First, even in collinear limit we found that $A^{0 0 - T_{2g}}_{1}= -0.53$ mRy while $A^{0 0 - E_{g}}_{1}= 0.12$ mRy, while $A^{zz - T_{2g}}_{1}=0.46$ mRy compared to the larger $E_{g}$ and mixed contributions, -0.45 mRy, and -1.1 mRy, respectively. The source of the anisotropic term (which was reported already in Ref. \cite{lounis}) is induced by the spin-polarized host due to double exchange mechanism. Note that in a paramagnetic phase the $A^{\mu\mu}_{ij}$ terms ($\mu=x,y,z$) are the same in statistical sense, i.e., we cannot speak of terms that are induced by the symmetry broken host \cite{szilva}. The role of the host induced terms can be crucial and should be analyzed at finite temperature studies, well above the ordering temperature, when the ferromagnetic background is vanishing, as well as their contribution to the Dzyaloshinskii-Moriya (DM) interaction which appears when the spin-orbit interaction is present and the inversion symmetry between the spins is lacking in the system.

We now show that the role of the host induced terms is more significant for non-collinear (non-equilibrium) systems by considering a relatively simple \textit{single spin rotation} of a non-collinear system (see Fig. \ref{J1}). In Fig. \ref{J1} the black line shows the relationship between the non-collinear Heisenberg and non-Heisenberg interatomic exchange coupling constants, see Eqs. (\ref{H}) and (\ref{NH}), for a nearest neighbour pair in bcc Fe. The parameters were calculated self-consistently for different $\theta_{i}$ angles (see dots) starting from the collinear ferromagnetic case where $J^{H}_{1}=0.85$ mRy when $\theta_{i}$ was 0, marked by the bigger black symbol. Note that in the final state $\theta_{i}=\pi$ which is a collinear state. In this case $J^{NH}_{1}=0$ and $J^{H}_{1}$ can be calculated in terms of LKAG formalism. The biggest relative change going from 0 to $\pi$ takes place in the $E_{g}$ channel (red line), i.e. this channel together with the mixed contribution is responsible for the configuration dependence of $J^{H}_{1}$ as well as the emergence of $J^{NH}_{1}$. However, the $J^{H}_{1}$ contribution in the $T_{2g}$ channel (see the data with blue) hardly changes as a function of $\theta_{i}$. In addition, the $J^{NH}_{1}$ contribution in the $T_{2g}$ remains small compared to the other channels. This clearly shows for the $T_{2}$ subspace that the mapping onto a Heisenberg spin Hamiltonian can be done even in the case when one spin is forced to be rotated by a large angle in a ferromagnetic background.

In Fig. \ref{J2}, where we show the corresponding plot for second nearest neighbor interactions, the obtained parameters are in the same energy range in all symmetry channels. Note that the magnitude of $J_{2}^{H}$ and $J_{2}^{NH}$ is much less than of $J_{1}^{H}$ and $J_{1}^{NH}$. It can be seen in Fig. \ref{NN} where the $J^{H}$ and $J^{NH}$ parameters are shown not only for the first and second but also for a third, fourth, fifth and sixth neighbouring pairs in bcc Fe. Within an agreement with the previous finding \cite{oldlicht, szilva, yaro, frota}, the first and second neighbor $J_{ij}$'s dominate the $J_{ij}$'s calculated for further neighbors. This conclusion holds in the non-collinear framework too. Finally, Figs \ref{NNE}, \ref{NNT2} and \ref{NNmix} show the results for the first six nearest neighbors $J^{H}$ and $J^{NH}$ in the $E_{g}$, $T_{2g}$ and mixed symmetry channels, respectively. Interestingly, one can see that in the $T_{2g}$ case (and only this case) shown in Fig. \ref{NNT2}, the first and second neighbor contribution (black and blue data) are in the same energy range.






\section{V. Conclusions}

To calculate experimentally detectable quantities as the critical temperature and the magnon excitation spectra, it is convenient to perform a spin dynamics simulations {\cite{ASDbook} with the description of motion of the atomistic spin. For this purpose the calculation of the interaction parameters between the spins is usually needed.

We have derived for a simple non-collinear spin configuration (when one spin was rotated by a finite angle in a ferromagnetic background) an equation for the energy of the pairwise interaction with terms induced by the spin-polarized host. These terms cannot be completely described by a bilinear, Heisenberg spin model. Similar but higher order host-induced terms were previously found even in collinear limit also in the lack of spin-orbit coupling when the anisotropic type terms were mapped onto higher order (biquadratic) spin Hamiltonian \cite{lounis, szilva}. 

We have shown, however, that an $E_{g}$-$T_{2g}$ symmetry analysis based on the decomposition of the $J_{ij}$ for the different atomic orbitals in bcc Fe leads to the conclusion that the nearest neighbor exchange parameters in the $T_{2g}$ channel are $essentially$ Heisenberg parameters. This means that they do not depend strongly on the spin configuration \textit{and} can be exactly mapped onto a Heisenberg spin model. These findings are in a very good agreement with the conclusions of Ref. \cite{yaro} based on the LKAG formalism. We also note that the nearest neighbour angles between the atomistic spins in bcc Fe are usually small at low temperatures (where the background is strongly spin polarized), hence a Heisenberg model with the LKAG exchange formula can give a good approximation. While at high temperature, when the angles between the neighbouring spins are larger, the background is less polarized.

The role of the host induced terms and their microscopic origin are crucial (and should be analyzed further) at finite temperature systems when the ferromagnetic background is vanishing as well as their contribution to the Dzyaloshinskii-Moriya interactions in relativistic calculations. These findings have key importance for strong out-of-equilibrium situations, and motivate a general description for equilibrium case that holds from the low temperature (LKAG range) up to the critical temperature (paramagnetic phase). The results can also motivate to study the non-Heisenberg behaviour of experimentally realistic systems as ultrafast demagnetization with the use of direct effective-field calculation.

\section{Acknowledgement}

We owe thanks to L. Szunyogh, L. Udvardi, A. De\'ak, M. Katsnelson, A. Liechtenstein for the fruitful discussions. Supports from the Swedish Research Council (VR), the KAW foundation (grants 2012.0031 and 2013.0020) and eSSENCE are acknowledged.The computations were performed on resources provided by the Swedish National Infrastructure for Computing (SNIC). A. Bergman acknowledges support from CEA-Enhanced Eurotalents, co-funded by FP7 Marie Sk\l{}odowska-Curie COFUND Programme (Grant Agreement n° 600382). P. F. Bessarab acknowledges support from the Icelandic Research Fund (Grant No. 163048-052) and the mega-grant of the Ministry of Education and Science of the Russian Federation (grant no. 14.Y26.31.0015). R. Cardias, D. C. M. Rodrigues, and A. B. Klautau acknowledge financial
support from CAPES and CNPq, Brazil.

\end{document}